\documentclass[aps,pra,twoside,10pt,floatfix,nofootinbib,
               twocolumn]{revtex4-1}

\usepackage[colorlinks=true, linkcolor=blue, citecolor=blue, urlcolor=blue]{hyperref}
\usepackage{amssymb,amsfonts,amsmath,bm,amsthm,mathtools,
            braket,xcolor,comment}
\usepackage[normalem]{ulem}
\usepackage{graphicx}
\usepackage{subcaption}
\usepackage{makecell}
\usepackage{booktabs}

\begin{document}


\title{\textbf{Scaling of Quantum Resources for Simulating a Long-Range System}}

\author{Tanya Keshari$^1$, Debasis Sadhukhan$^{2,3}$}
\affiliation{$^1$ Department of Physics, Indian Institute of Technology Roorkee, Roorkee, 247 667, Uttarakhand, India \\
$^2$ Department of Physics, Institute of Science, Banaras Hindu University, Varanasi, 221 005, Uttar Pradesh, India\\
$^3$ Department of Physics, Indian Institute of Engineering Science and Technology, Shibpur, 711 103, West Bengal, India}

\begin{abstract}

We simulate a long-range extended Ising model in one dimension using a hybrid quantum algorithm, namely Variational Quantum Eigensolver (VQE). In this quantum simulation, we investigate how quantum resources scale with system size and interaction strength. Three structure-aware ans\"{a}tze incorporating nearest-neighbor (NN), 
next-nearest-neighbor (NNN), and next-next-nearest-neighbor 
(NNNN) entangling blocks are constructed by mimicking the string operators in the Hamiltonian. We show that energy 
fidelity alone is not a good indicator for finding the ground state of our model. To overcome this problem, we introduce an additional criterion based on pairwise logarithmic negativity 
as a more reliable way to find the actual ground state by the VQE. We find that the 
interaction range parameter $\alpha$ primarily governs the 
minimum number of ansatz layers required, rather than proximity 
to the quantum critical point. In particular, we show that in the non-local regime \(\alpha \leq 1\), the NNN and NNNN ans\"{a}tze reduce the layer scaling rate by factors of $2.5\times$ and $3.8\times$ 
relative to NN in all phases, including the critical point. The total number of two-qubit gates required for reliable simulation grows quadratically with system size for all three ansatzes. This is consistent with the theoretical prediction, as the number of non-local terms in the Hamiltonian also grows quadratically with the system size. In the local regime, however, the number of required two-qubit gates grows linearly with system size. In contrast, in the quasi-local regime, the required number of two-qubit gates for the quantum simulation is more subtle and depends on the phase of the Hamiltonian.

\end{abstract}

\maketitle

\section{Introduction}
\label{sec:intro}
Efficient simulation of quantum many-body systems is one of the central promises in quantum computing. The core difficulty is that the number of quantum states grows exponentially with the number of particles. For a chain of $N$ spin-$\frac{1}{2}$ particles, the Hilbert space has $2^N$ dimensions, growing so rapidly that exact diagonalization becomes impossible beyond roughly $30$--$40$ sites on the most powerful classical computers available today. Tensor-network methods such as the Density Matrix Renormalization Group (DMRG) \cite{ref1, ref2} offer a way around this for many systems. They work by exploiting the fact that ground states of local Hamiltonians typically do not use their full Hilbert space; instead, the entanglement between any region and its surroundings scales with the boundary area rather than the volume, a property known as the area law \cite{ref3, ref4}. In one-dimensional gapped systems, this area law holds rigorously \cite{ref4}, and even at quantum critical points, the entanglement entropy grows only logarithmically \cite{ref5}. However, when the area law is violated, tensor network methods require exponentially large bond dimensions, severely hindering their usability.

Long-range interacting systems are one such example where this breakdown occurs. When the coupling between spins decays algebraically as $J_r \sim r^{-\alpha}$, the ground state can carry correlations that span the entire system, and the entanglement entropy can grow logarithmically or even faster depending on $\alpha$ \cite{ref6, ref7, ref8}. The physics changes qualitatively across three regimes: long-range ($\alpha < 1$), quasi-local ($1 < \alpha < 2$), and effectively short-range ($\alpha \gg 2$) \cite{ref9}. Even the way information spreads through these systems is fundamentally different. While local systems obey a strict Lieb-Robinson light cone that limits how fast correlations can propagate, long-range systems can transmit correlations much faster, and for small enough $\alpha$, this propagation violates generalized locality bounds entirely \cite{ref10}. Such long-range interactions arise naturally in trapped-ion quantum simulators \cite{ref11, ref12} and Rydberg atom arrays \cite{ref13}, where the exponent $\alpha$ can be tuned experimentally, making these systems ideal testbeds for exploring the crossover between interaction regimes. Despite this experimental accessibility, efficiently simulating long-range quantum spin models on near-term quantum hardware remains an open problem.

The arrival of Noisy Intermediate-Scale Quantum (NISQ) devices, quantum processors with tens to hundreds of qubits that operate without full error correction \cite{ref14}, has opened a realistic near-term route toward simulating classically intractable systems. The most promising algorithm for this purpose is the Variational Quantum Eigensolver (VQE) \cite{ref15, ref16}. The idea is simple: a parameterized quantum circuit, called the ansatz, prepares a trial quantum state, and a classical computer adjusts the circuit parameters to minimize the energy expectation value of the target Hamiltonian. This hybrid quantum-classical approach keeps the quantum circuit relatively shallow, which is important because deeper circuits accumulate more noise from imperfect 
gates \cite{ref17, ref18}. VQE is well matched to NISQ hardware, where two-qubit gate error rates are still significant \cite{ref19}, coherence times are short \cite{ref14, ref20}, and experimental demonstrations have already shown its viability for quantum chemistry and lattice model simulation \cite{ref21}.

The practical bottleneck in VQE is the choice of ansatz. On the one hand, the ansatz must be expressive enough to represent the target ground state well. On the other hand, it must be shallow enough to run on noisy hardware without accumulating too many errors. Generic hardware-efficient ans\"{a}tze, which are designed primarily to fit the connectivity of the device \cite{ref22}, often lack the expressibility needed for systems with complex entanglement. Worse, when such circuits are initialized randomly at large depth, they suffer from the barren plateau problem: the gradient of the cost function becomes exponentially small with the number of qubits, making it practically impossible for the optimizer to find its way to the ground state \cite{ref23}. A better approach is to design ans\"{a}tze that are inspired by the physics of the target Hamiltonian, encoding the relevant entanglement structure directly into the circuit. Such problem-inspired ans\"{a}tze achieve higher accuracy at shallower depths and maintain trainable optimization landscapes \cite{ref24, ref25}. The expressibility and entangling capability of parameterized circuits have been studied systematically \cite{ref26}, and the link between the entanglement structure of the circuit and its optimization performance has been established for specific Hamiltonians \cite{ref27}. Comparisons of resource costs across different variational strategies confirm that what matters is not just the depth of the circuit but its structural alignment with the target state \cite{ref28}. Despite all of this progress, no study has systematically and quantitatively examined how the entanglement range of the ansatz must be matched to the interaction range of the Hamiltonian, and how the resulting resource costs scale with system size and interaction regime.

A closely related question was recently studied in the classical setting by Lakkaraju \textit{et al.} \cite{ref29}. They studied the one-dimensional long-range extended Ising model and asked whether the pairwise entanglement of its ground state, quantified by the logarithmic negativity between any two sites \cite{ref30, ref31}, could be reproduced by a Hamiltonian with only a finite number of neighboring interactions. Their answer depends critically on $\alpha$. In the quasi-local regime ($1 < \alpha < 2$), only about $O(10)$ neighboring interactions are enough to reproduce the entanglement pattern of the fully connected model. But in the nonlocal regime ($\alpha < 1$), entanglement develops an algebraic tail with distance, scaling roughly as $E_r \sim r^{-\alpha}$, and no finite number of interactions suffices; you need approximately $O(N)$ interactions to mimic the true long-range model. The monogamy score of entanglement \cite{ref32, ref33} was found to be a reliable indicator of this transition: it saturates with increasing interaction range when finite-range mimicry is achievable, and decays algebraically when it is not. This result immediately raises a parallel question for quantum simulation: if a VQE ansatz uses only finite-range entangling operations, can it faithfully reproduce the ground-state entanglement of the long-range model and at what cost in terms of circuit depth and computational resources?

The long-range extended Ising model is an ideal setting for this investigation because it can be solved exactly. Its interactions are mediated by string operators of the form $\sigma^x_n \prod_{l=n+1}^{n+r-1} \sigma^z_l \, \sigma^x_{n+r}$, which are constructed so that the non-local phases from the Jordan-Wigner transformation cancel exactly, making the model analytically tractable through a mapping to free fermions \cite{ref34}. This means we can use full exact diagonalization as a benchmark; any difference in VQE performance between ans\"{a}tze is a genuine physical effect, not a numerical artifact. The model also has a rich phase diagram with paramagnetic, critical, and ferromagnetic phases, and its non-local entanglement structure makes it a representative challenge for the broader class of long-range interacting systems.

In this work, we apply VQE to this model across three interaction regimes ($\alpha = 0.5$, $1.5$, $10.0$), three coupling ratios covering the paramagnetic, critical, and ferromagnetic phases ($|J/h| = 0.5$, $1.0$, $2.0$), and system sizes $N = 4$ to $9$. We design three structure-aware ans\"{a}tze incorporating nearest-neighbor (NN), next-nearest-neighbor (NNN), and next-next-nearest-neighbor (NNNN) multipartite entangling blocks that directly mirror the string operators in the Hamiltonian and evaluate performance using pairwise logarithmic negativity \cite{ref30, ref31} as the primary success criterion rather than energy fidelity alone. The logarithmic negativity is computed via partial transpose of the bipartite reduced density matrix \cite{ref35}. We show that states with fidelity above $0.99$ can still fail to reproduce the correct long-range entanglement, confirming that energy convergence alone is not a sufficient test of VQE success in long-range systems. By tracking the minimal number of ansatz layers $p^*$, the total CNOT gate cost $R_Q$, and the classical optimization cost $R_C$ needed to meet the entanglement-based criterion, we establish a direct and quantitative connection between the interaction range of the Hamiltonian and the entanglement range required of the ansatz.

Our main findings are the following. In the long-range regime ($\alpha = 0.5$), adding three-body NNN and four-body NNNN entangling blocks that mirror the Hamiltonian's string operators reduces the layer scaling rate by factors of $2.6$ and $3.8$ relative to NN for the NNN and NNNN ans\"{a}tze respectively, with this advantage strengthening as the system grows larger. Both quantum and classical resource costs scale as $O(N^2)$ in this regime, and the layer reduction translates into proportional or greater resource savings. In the quasi-local regime ($\alpha = 1.5$), the advantage of the multipartite ans\"{a}tze is strongest near the quantum critical point and fades in the ferromagnetic phase. In the short-range regime ($\alpha = 10.0$), all three ans\"{a}tze perform similarly, and the simpler NN circuit is the most efficient choice. We also find that the interaction-range parameter $\alpha$ is not the proximity to the quantum critical point but primarily determines the layer requirements, a physically non-trivial result with direct implications for ansatz design.

The rest of the paper is organized as follows. Section~\ref{sec:model} introduces the long-range extended Ising model and its exact solution via the Jordan-Wigner transformation. Section~\ref{sec:method} describes the VQE framework, the three ans\"{a}tze, and the entanglement-based success criterion. Section~\ref{sec:results} presents the results for minimal 
ansatz layers, quantum resource cost, and classical resource 
cost, along with their physical interpretation. 
Section~\ref{sec:conclusion} concludes and discusses 
limitations and directions for future work.

\section{The Long-Range Extended Ising Model}
\label{sec:model}
We consider a one-dimensional spin-$\frac{1}{2}$ chain of $N$ sites described by a modified long-range transverse-field Ising model \cite{ref29, ref34}. The Hamiltonian is:

\begin{equation}
\hat{H} = \sum_{n=1}^{N}\left[ h \sigma_n^z - \sum_{r=1}^{N-1} J_r \left( \sigma_n^x \prod_{l=n+1}^{n+r-1} \sigma_l^z \, \sigma_{n+r}^x \right)\right],
\label{eq:hamiltonian}
\end{equation}
where $\sigma^x$ and $\sigma^z$ are the Pauli matrices, $h$ is the strength of the transverse magnetic field, and $J_r$ is the coupling strength between spins separated by distance $r$.

A key feature of this model is the $\sigma^z$-string operator $\prod_{l=n+1}^{n+r-1}\sigma_l^z$ connecting each interacting pair. In a standard long-range $\sigma^x\sigma^x$ model, applying the Jordan-Wigner transformation introduces non-local string operators that prevent an exact analytical treatment. The $\sigma^z$-string in our Hamiltonian is specifically constructed so that these non-local phases cancel exactly upon applying the transformation, yielding a purely quadratic fermionic Hamiltonian that can be solved analytically \cite{ref29, ref34}.

\subsection{Diagonalization}

The spin degrees of freedom are mapped to spinless fermions via the Jordan-Wigner transformation \cite{ref34}:
\begin{equation}
\sigma_n^z = 2c_n^\dagger c_n - 1, \qquad \sigma_n^x = (c_n^\dagger + c_n)\prod_{m<n}(1 - 2c_m^\dagger c_m),
\end{equation}
where the fermionic operators satisfy $\{c_m, c_n^\dagger\} = \delta_{mn}$ and $\{c_m, c_n\} = 0$. The cancellation of the non-local JW phases by the $\sigma^z$-string yields a simple quadratic interaction term:

\begin{equation}
\sigma_n^x \prod_{l=n+1}^{n+r-1}\sigma_l^z \, \sigma_{n+r}^x = (c_n^\dagger - c_n)(c_{n+r}^\dagger + c_{n+r}).
\end{equation}
Substituting into Eq.~(\ref{eq:hamiltonian}), the Hamiltonian in real space becomes:
\begin{multline}
\hat{H} = -\sum_{n=1}^{N}\sum_{r=1}^{N-n} J_r\left(c_n^\dagger c_{n+r} 
+ c_n^\dagger c_{n+r}^\dagger - c_n c_{n+r} - c_n c_{n+r}^\dagger\right) \\
+ 2h\sum_{n=1}^{N}c_n^\dagger c_n - hN.
\end{multline}
This quadratic structure allows the system to be diagonalized exactly into non-interacting quasiparticles. Transforming to momentum space via a Fourier transform followed by a Bogoliubov transformation, the Hamiltonian takes the Bogoliubov-de Gennes (BdG) form:
\begin{equation}
\hat{H} = \sum_{k>0}
\begin{pmatrix}
c_k^\dagger & c_{-k}
\end{pmatrix}
\begin{pmatrix}
\epsilon_k & i\Delta_k \\
-i\Delta_k & -\epsilon_k
\end{pmatrix}
\begin{pmatrix}
c_k \\
c_{-k}^\dagger
\end{pmatrix}
- hN,
\end{equation}
with quasiparticle dispersion:
\begin{equation}
E(k) = \sqrt{\epsilon_k^2 + \Delta_k^2},
\end{equation}
where:
\begin{align}
\epsilon_k &= h - \sum_{r=1}^{N-1} J_r \cos(kr), \\
\Delta_k &= \sum_{r=1}^{N-1} J_r \sin(kr).
\end{align}
A quantum phase transition occurs when the energy gap $E(k)$ vanishes. At $k = 0$ and $\pi$, this gives the critical field $h_c = \sum_{r=1}^{N-1} J_r$ and $\sum_{r=1}^{N-1} (-1)^r J_r$ respectively. However, in this work we primarily concentrate on the $k=0$ case.

\subsection{Parameter Scaling and Interaction Regimes}

The long-range coupling is defined as:
\begin{equation}
J_r = \frac{J}{A} \, r^{-\alpha},
\end{equation}
where $J$ is the overall interaction energy scale, $\alpha$ is the exponent governing the decay of interactions with distance, and $A$ is a normalization constant. To ensure that the total energy remains extensive in the thermodynamic limit, a requirement known as Kac scaling \cite{ref37, ref9}, $A$ is chosen as:

\begin{equation}
A = \sum_{r=1}^{N-1} r^{-\alpha}.
\end{equation}
We define the dimensionless coupling ratio $\lambda = J/h$ to characterize the competition between the interactions and the transverse field. Setting $h = A$ fixes the critical condition $h_c = \sum_{r=1}^{N-1} J_r = J$ at $\lambda = 1$, independent of $N$ and $\alpha$. The system is therefore in the paramagnetic phase for $\lambda < 1$, at the quantum critical point at $\lambda = 1$, and in the ferromagnetic phase for $\lambda > 1$.

The exponent $\alpha$ controls the physical character of the interactions \cite{ref9, ref29}. For $0 \leq \alpha < 1$, interactions are truly long-range and correlations span the entire system. For $1 < \alpha < 2$, interactions are quasi-local with intermediate correlation lengths. For $\alpha \gg 2$, the model effectively recovers nearest-neighbor Ising behavior. In this work we study three representative values: $\alpha = 0.5$ (long-range), $\alpha = 1.5$ (quasi-local), and $\alpha = 10.0$ (effectively short-range), at three coupling ratios $\lambda = 0.5$, $1.0$, and $2.0$, spanning the paramagnetic, critical, and ferromagnetic phases respectively.

\section{Method}
\label{sec:method}
\subsection{Variational Quantum Eigensolver}

The Variational Quantum Eigensolver (VQE) \cite{ref15} is currently one of the most practical approaches for computing the ground state of a quantum Hamiltonian on NISQ hardware. It is based on the variational principle: for any normalized trial wavefunction $|\psi\rangle$, the expectation value $\langle\psi|\hat{H}|\psi\rangle$ is always greater than or equal to the true ground-state energy $E_0$. The ground state can therefore be approximated by finding the trial state that minimizes this expectation value.

\begin{figure*}[t] 
    \centering
    \vspace{-7em} 

    \begin{subfigure}[b]{0.7\linewidth}
        \centering
        \includegraphics[width=\linewidth]{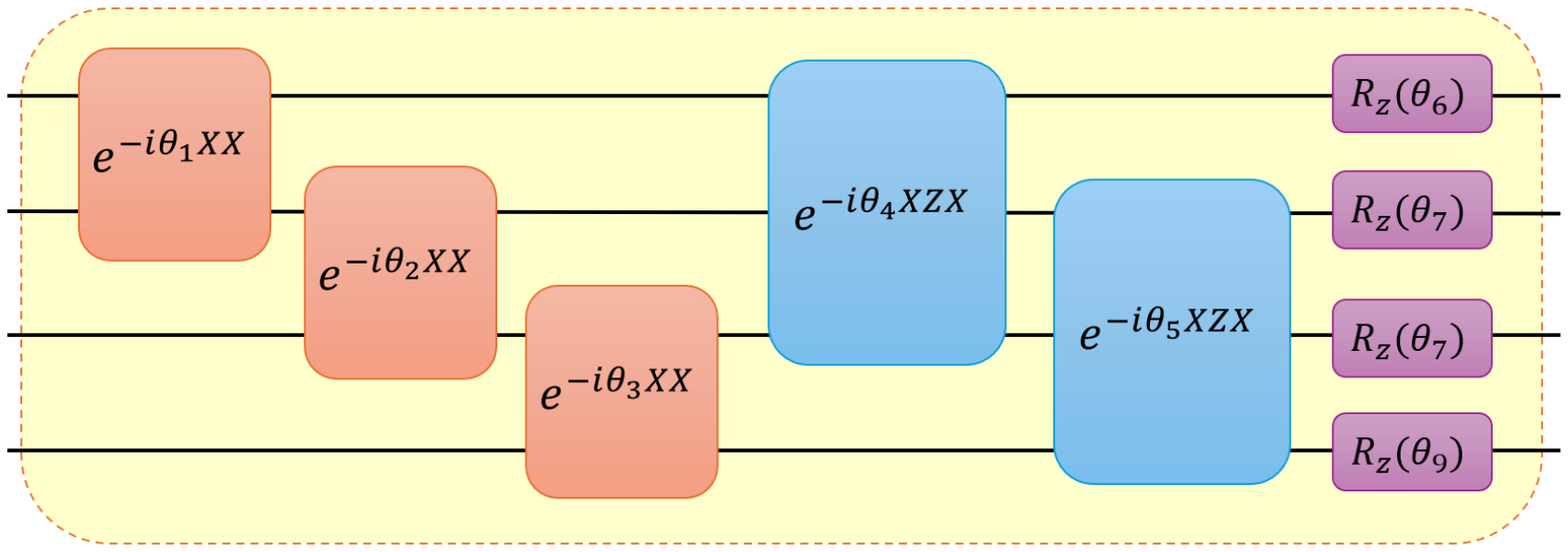}
        \label{fig:circuit1}
    \end{subfigure}
    
    \vspace{-10em} 

    \begin{subfigure}[b]{0.45\linewidth}
        \centering
        \includegraphics[width=\linewidth]{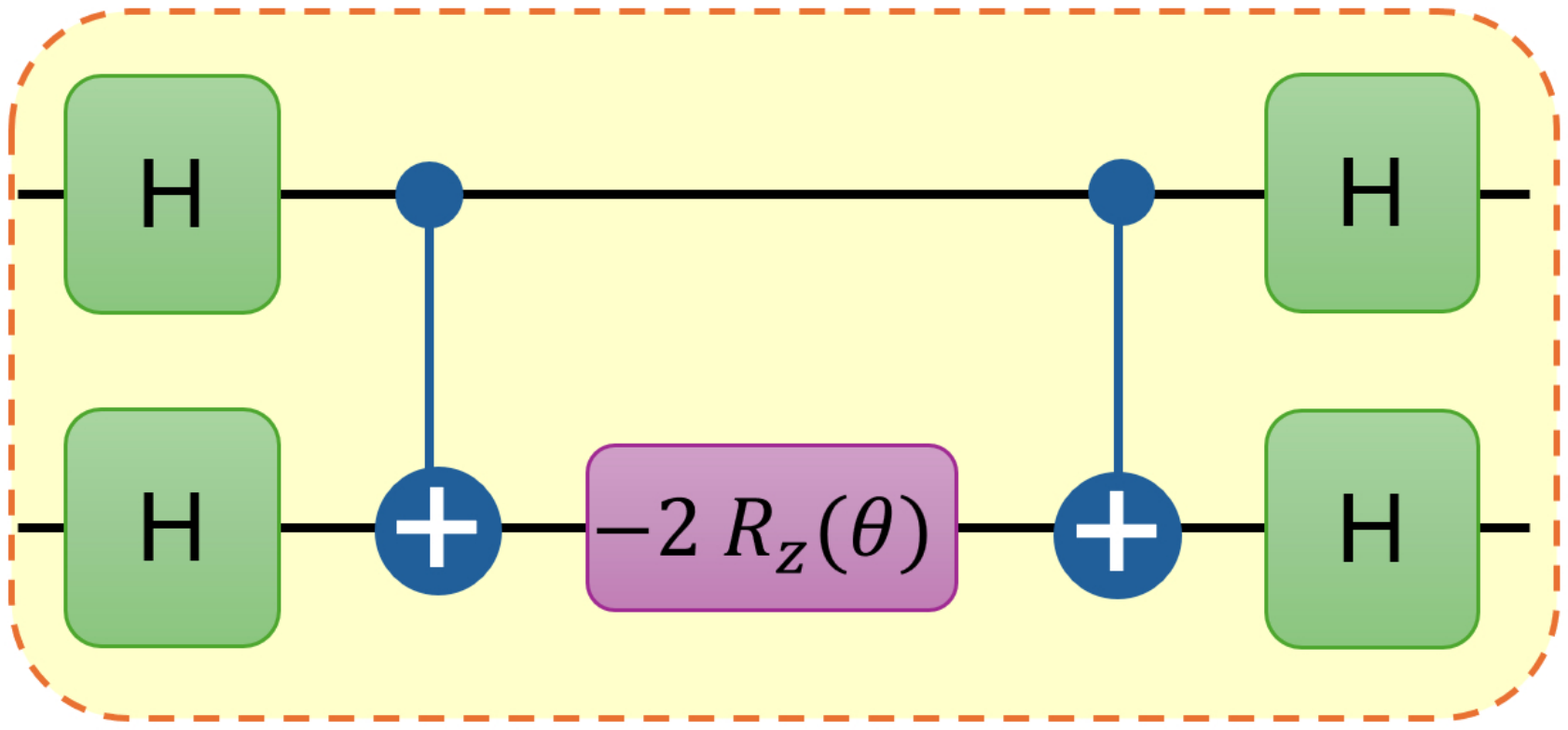}
        \label{fig:circuit2}
    \end{subfigure}
    \hfill 
    \begin{subfigure}[b]{0.43\linewidth}
        \centering
        \includegraphics[width=\linewidth]{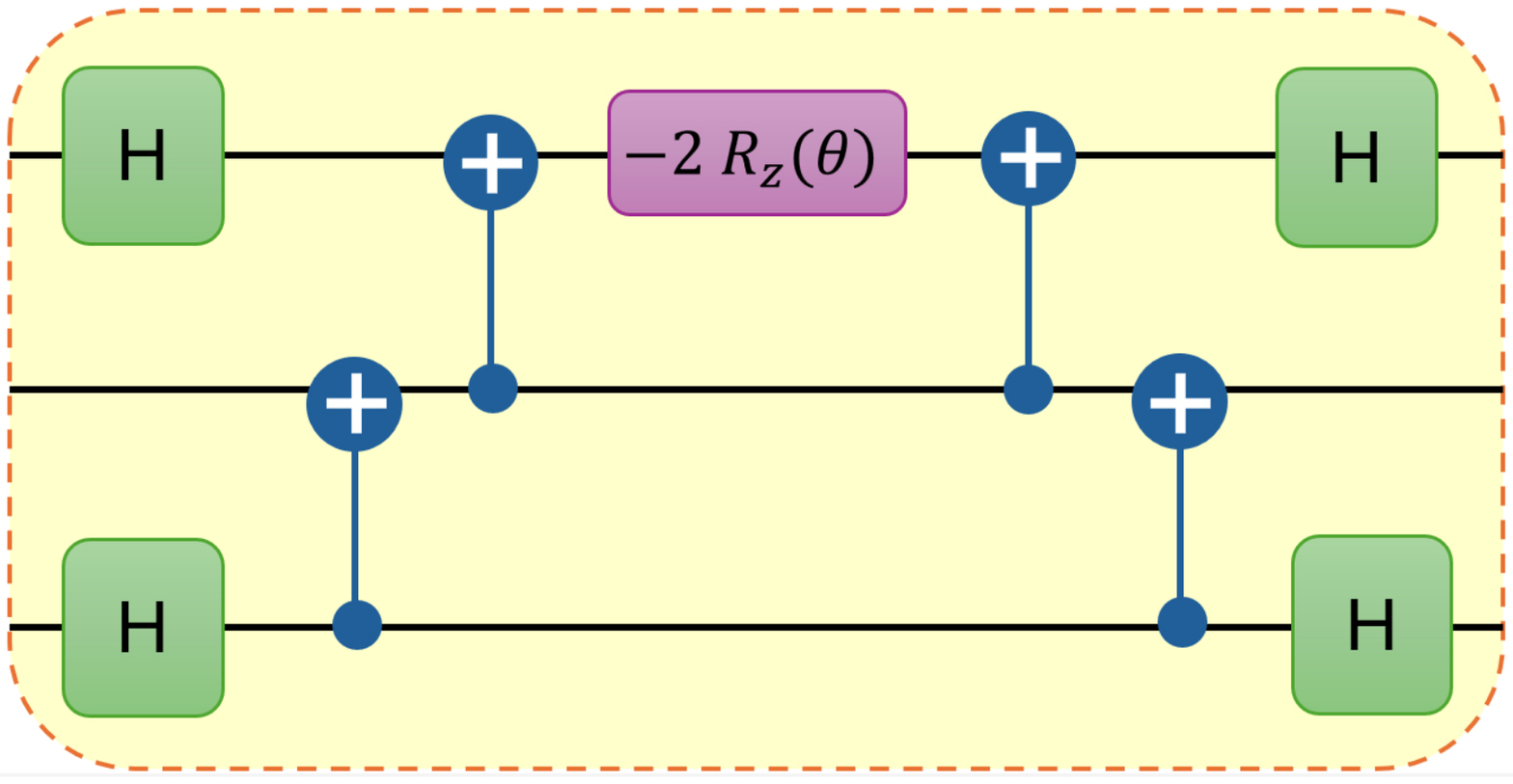}
        \label{fig:circuit3}
    \end{subfigure}

    \vspace{-2em}
    
    \caption{(Top panel) A single layer of the Next-Nearest-Neighbor (NNN) ansatz for system size $N = 4$. The circuit consists of entangling blocks implementing the unitary evolutions $e^{-i\theta \sigma_j^x \sigma_{j+1}^x}$ (NN) and $e^{-i\theta \sigma_j^x \sigma_{j+1}^z \sigma_{j+2}^x}$ (NNN), followed by local $R_z(\theta)$ rotations on each qubit. (Bottom left panel) and (Bottom right panel) show the 
    decomposition of the two-qubit $e^{-i\theta XX}$ and three-qubit 
    $e^{-i\theta XZX}$ entangling blocks respectively into the native gate set 
    $\{H, \text{CNOT}, R_z\}$. The four-qubit $e^{-i\theta XZZX}$ (NNNN) block 
    follows the same staircase pattern as the three-qubit block, with two additional CNOTs (6 total).}
    \label{fig:circuit}
\end{figure*}

In practice, VQE operates as a hybrid loop between a quantum processor and a classical computer. On the quantum side, a parameterized circuit called the ansatz starts from an initial state $|0\rangle^{\otimes N}$ and applies a sequence of unitary operations controlled by a set of angles $\boldsymbol{\theta}$, producing the trial state $|\psi(\boldsymbol{\theta})\rangle = U(\boldsymbol{\theta})|0\rangle^{\otimes N}$. The expectation value of the Hamiltonian,
\begin{equation}
E(\boldsymbol{\theta}) = \langle\psi(\boldsymbol{\theta})|\hat{H}|\psi(\boldsymbol{\theta})\rangle,
\end{equation}
serves as the cost function. This value is passed to a classical optimizer; in this work, the L-BFGS-B algorithm \cite{ref39} updates $\boldsymbol{\theta}$ to reduce the cost. The cycle of state preparation, measurement, and parameter update repeats until convergence, yielding the approximate ground-state energy $E_{\text{VQE}} \approx E_0$ and the corresponding state $|\psi(\boldsymbol{\theta}^*)\rangle$.

VQE is well suited to NISQ devices because it can operate with relatively shallow circuits. Deep circuits accumulate too much noise from imperfect gates, particularly two-qubit gates which are significantly noisier than single-qubit operations \cite{ref19, ref20}. By shifting part of the computational burden to classical optimization, VQE provides a practical compromise for devices where coherence times are short and gate error rates remain significant.

The practical success of VQE depends heavily on efficient use of both quantum and classical resources. The quantum resource cost $R_Q$ is measured by the total number of two-qubit CNOT gates in the full parameterized circuit \cite{ref26, ref27, ref28}. The classical resource cost captures the total optimization burden and is quantified as:
\begin{equation}
R_C = p^* \times L \times n_I,
\end{equation}
where $p^*$ is the minimal number of ansatz layers, $L$ is the number of variational parameters per layer, and $n_I$ is the average number of optimizer iterations required to reach convergence \cite{ref28, ref48}. Reducing both $R_Q$ and $R_C$ is essential for making VQE scalable on near-term hardware.

\subsection{Variational Ansatz}

Choosing the right ansatz is one of the most important decisions in any VQE calculation. Its structure directly affects how accurately the ground state can be approximated, how quickly the optimizer converges, and how many quantum gates are required on NISQ hardware.

Generic hardware-efficient ans\"{a}tze \cite{ref22} are designed for device connectivity rather than physical structure, and we found them insufficient for reliably reproducing the long-range entanglement profiles of this model. We therefore designed three structure-aware ans\"{a}tze whose entangling blocks directly mirror the string operators in the Hamiltonian.

Long-range Hamiltonians require the quantum state to establish correlations across many sites, and standard nearest-neighbor gates are inefficient at generating these. Directly simulating a fully connected long-range model would require at least $O(N!)$ two-qubit gates for a system of $N$ sites \cite{ref29}, making NISQ-era simulations extremely noisy. The aim here is to determine whether structure-aware entangling gates can capture the key physics while using substantially fewer resources.

The custom ansatz is shown in Fig.~\ref{fig:circuit}. The circuit is assembled layer by layer, where each layer consists of entangling blocks followed by single-qubit $R_z(\theta)$ rotations on each qubit. The $R_z$ rotations provide local phase control and improve the ansatz's ability to refine the state without substantially increasing circuit depth. We consider three ansatz configurations to study how the entanglement range of the circuit affects performance:

\begin{itemize}
    \item \textbf{Nearest-neighbor (NN):} Each entangling block implements the parametrized evolution $e^{-i\theta\sigma_j^x\sigma_{j+1}^x}$. This is the simplest case, analogous to a standard XX model, and serves as the reference ansatz.

    \item \textbf{Next-nearest-neighbor (NNN):} In addition to the NN blocks, three-site interactions are introduced via $e^{-i\theta\sigma_j^x\sigma_{j+1}^z\sigma_{j+2}^x}$. This directly mirrors the two-body string operators in the Hamiltonian.

    \item \textbf{Next-next-nearest-neighbor (NNNN):} Four-site interactions $e^{-i\theta\sigma_j^x\sigma_{j+1}^z\sigma_{j+2}^z\sigma_{j+3}^x}$ are further incorporated. This is the most expressive of the three configurations and most directly emulates the longer string operators present in the Hamiltonian.
\end{itemize}

The circuit is implemented in Qiskit \cite{Qiskit}. The gate-level decomposition of the two-qubit $e^{-i\theta XX}$ and three-qubit $e^{-i\theta XZX}$ entangling blocks into the native gate set $\{H, \text{CNOT}, R_z\}$ is shown in Figs.~\ref{fig:circuit}. The resource overhead per layer for each ansatz in terms of total CNOT gates and variational parameters is summarized in Table~\ref{tab:ansatz-resource-overhead}.

The choice of initial state was guided by the parity symmetry of the Jordan--Wigner-transformed model \cite{ref34}. For even $N$, the initial state is $|0\rangle^{\otimes N}$. For odd $N$, an $X$ gate is applied to each qubit, preparing $|1\rangle^{\otimes N}$. Since the ground state resides in the even-parity sector, initializing in the correct parity sector prevents the optimizer from exploring irrelevant regions of the Hilbert space.

\begin{table}[bp]
\centering
\setlength{\tabcolsep}{12pt}
\renewcommand{\arraystretch}{1.4}
\caption{Resource overhead per layer for the three ans\"{a}tze.}
\label{tab:ansatz-resource-overhead}
\begin{tabular}{lcc}
\toprule[1pt]
\textbf{Ansatz} & \makecell{\textbf{CNOT gates}\\\textbf{per layer}} & \makecell{\textbf{Parameters}\\\textbf{per layer ($L$)}} \\ 
\midrule[1pt]
NN    & $2(N-1)$   & $2N-1$    \\
NNN   & $2(3N-5)$  & $3(N-1)$  \\
NNNN  & $4(3N-7)$  & $2(2N-3)$ \\
\bottomrule[1pt]
\end{tabular}
\end{table}

\subsection{Ground State Success Criteria}
\label{Sec:GroundStateSuccessCriteria}

\begin{figure*}[t]
    \centering
    \includegraphics[width=\textwidth]{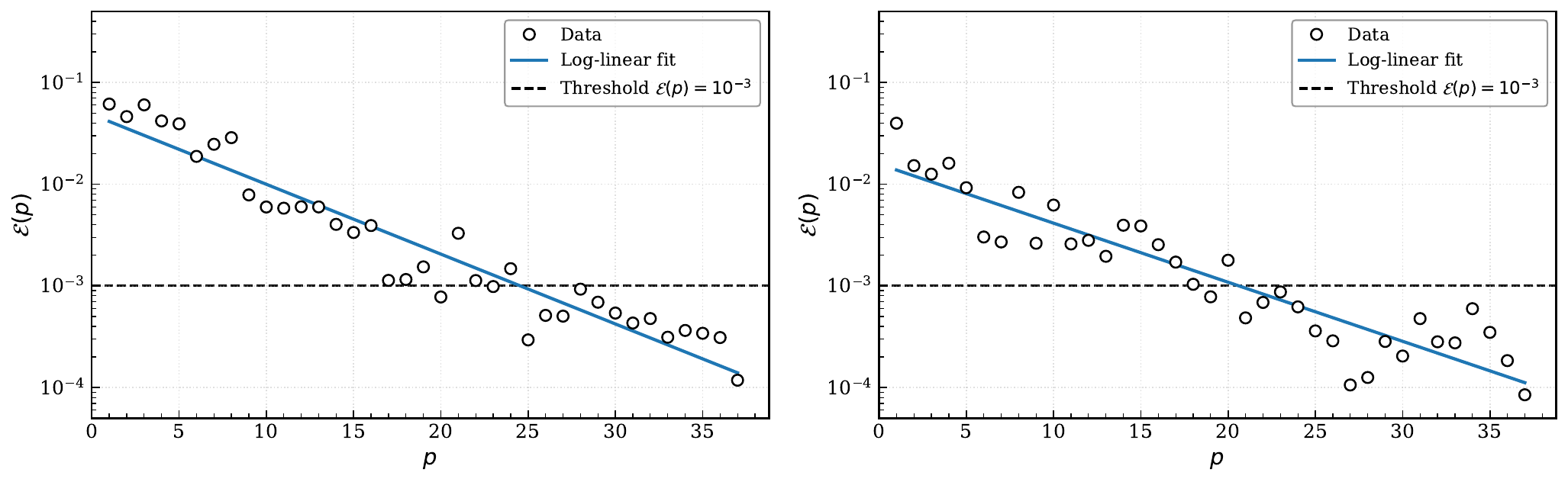}
    
    \caption{Total entanglement error $\mathcal{E}(p)$ as a function of ansatz layers $p$ on a log-linear scale for the NN ansatz, shown for two representative regimes at system size $N = 8$. (Left panel) Long-range regime ($\alpha = 0.5$, $\lambda = 0.5$). (Right panel) Quasi-local regime ($\alpha = 1.5$, $\lambda = 0.5$). Data points (open circles) are fitted with a log-linear model (solid line). The horizontal dashed line marks the convergence threshold $\mathcal{E}(p) = 10^{-3}$.}
    \label{fig:error_vs_p}
\end{figure*}

\begin{figure}[hbp]
    \centering
    \includegraphics[width=\linewidth]{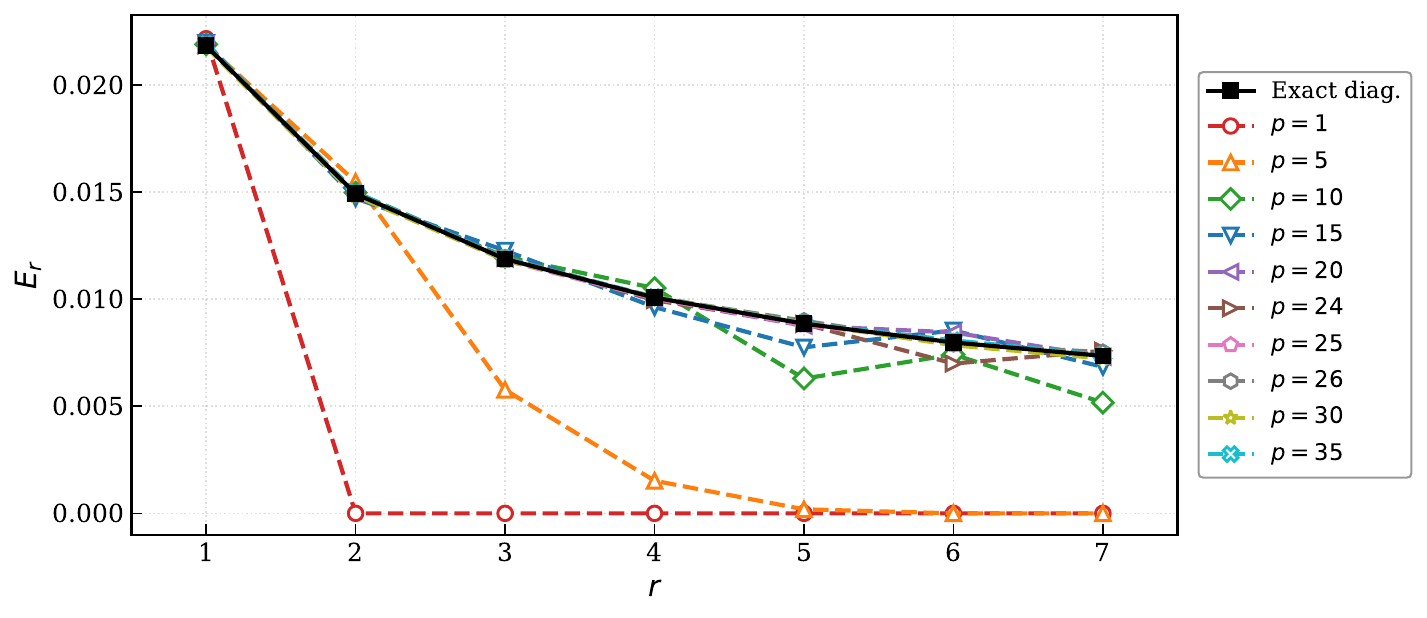}
    \caption{Pairwise logarithmic negativity $E_r$ as a function of distance $r$ for $N = 8$, $\alpha = 0.5$, and $\lambda = 0.5$ (NN ansatz). Exact 
    diagonalization (ED) results (solid line) are compared with VQE-prepared states at a selected value of ansatz layers $p$ (dashed lines). All shown VQE states achieve energy fidelity above $0.99$, yet only for $p \geq 25$ do the VQE profiles overlap with the ED result across all pairwise distances, confirming that energy fidelity alone is insufficient and that $\mathcal{E}(p) \leq 10^{-3}$ is a more reliable convergence threshold.}
    \label{fig:entanglement-profiles-NN}
\end{figure}

After VQE optimization, the quality of the converged state must be evaluated. A common criterion is a high state fidelity with the exact ground state, for example, fidelity greater than $0.99$. However, we found that this criterion alone was insufficient for long-range systems: states with fidelity above $0.99$ sometimes failed to reproduce the correct long-range entanglement profile, and accurate representation occasionally required fidelities as high as $0.999$ or even more. Energy fidelity is a global scalar that can be high even when the prepared state differs from the true ground state in its quantum correlations at longer distances. This is demonstrated explicitly in Fig.~\ref{fig:entanglement-profiles-NN}.

We therefore adopt pairwise logarithmic negativity \cite{ref30, ref31} as the primary success criterion. This quantity directly quantifies the strength of quantum correlations between any two sites and is sensitive to the algebraic decay of entanglement present in long-range systems. It is computed via the partial transpose of the bipartite reduced density matrix \cite{ref35}:
\begin{equation}
E_N(\rho_{AB}) = \log_2\|\rho_{AB}^{T_A}\|_1,
\end{equation}
where $\rho_{AB}^{T_A}$ denotes the partial transpose with respect to subsystem $A$, and $\|\rho\|_1 = \mathrm{tr}\sqrt{\rho^\dagger\rho}$ is the trace norm. Logarithmic negativity is chosen over alternative bipartite measures such as concurrence \cite{ref36} because it is directly computable without optimization for mixed states arising from any multipartite system.

To quantify how closely the VQE-prepared state reproduces the exact entanglement profile, we define the total absolute entanglement error across all pairwise distances as:
\begin{equation}
\mathcal{E}(p) = \sum_{r=1}^{N-1}\left|E_N^{\text{ED}}(r) - E_N^{\text{VQE}}(r, p)\right|,
\end{equation}
where $E_N^{\text{ED}}(r)$ is the exact logarithmic negativity from full diagonalization and $E_N^{\text{VQE}}(r, p)$ is the value from the VQE-prepared state for the number of ansatz layers $p$.

The minimal layer $p^*$ is defined as the smallest $p$ such that $\mathcal{E}(p) \leq 10^{-3}$. To determine this systematically across all parameter configurations, we fit the decay of $\mathcal{E}(p)$ with a log-linear model:
\begin{equation}
\ln\mathcal{E}(p) = ap + b,
\end{equation}
as shown in Fig.~\ref{fig:error_vs_p}. This form, corresponding to exponential decay of the entanglement error with number of ansatz layers, provided the best balance of fit quality, physical interpretability, and stability across all regimes and ans\"{a}tze.

The threshold of $10^{-3}$ was chosen empirically: it corresponds to visual agreement between the VQE and exact entanglement profiles, as demonstrated in Fig.~\ref{fig:entanglement-profiles-NN}. At this threshold, the VQE-prepared state not only achieves a high energy fidelity but also accurately reproduces the full entanglement structure of the true ground state across all pairwise distances.

\section{Results and Discussion}
\label{sec:results}
We present a systematic analysis of three quantities, the minimal number of ansatz layers $p^*$, the total quantum resource cost $R_Q$, and the total classical resource cost $R_C$ across the three ans\"{a}tze (NN, NNN, NNNN), three interaction regimes ($\alpha = 0.5$, $1.5$, $10.0$), three coupling ratios ($\lambda = 0.5$, $1.0$, $2.0$), and system sizes $N = 4$ to $9$. Together, these quantities characterize how efficiently each ansatz reproduces the ground-state entanglement structure of the long-range extended Ising model under NISQ-like constraints.

\subsection{Minimal Ansatz Layers}

\begin{figure*}[t]
    \centering
    \includegraphics[width=0.95\linewidth]{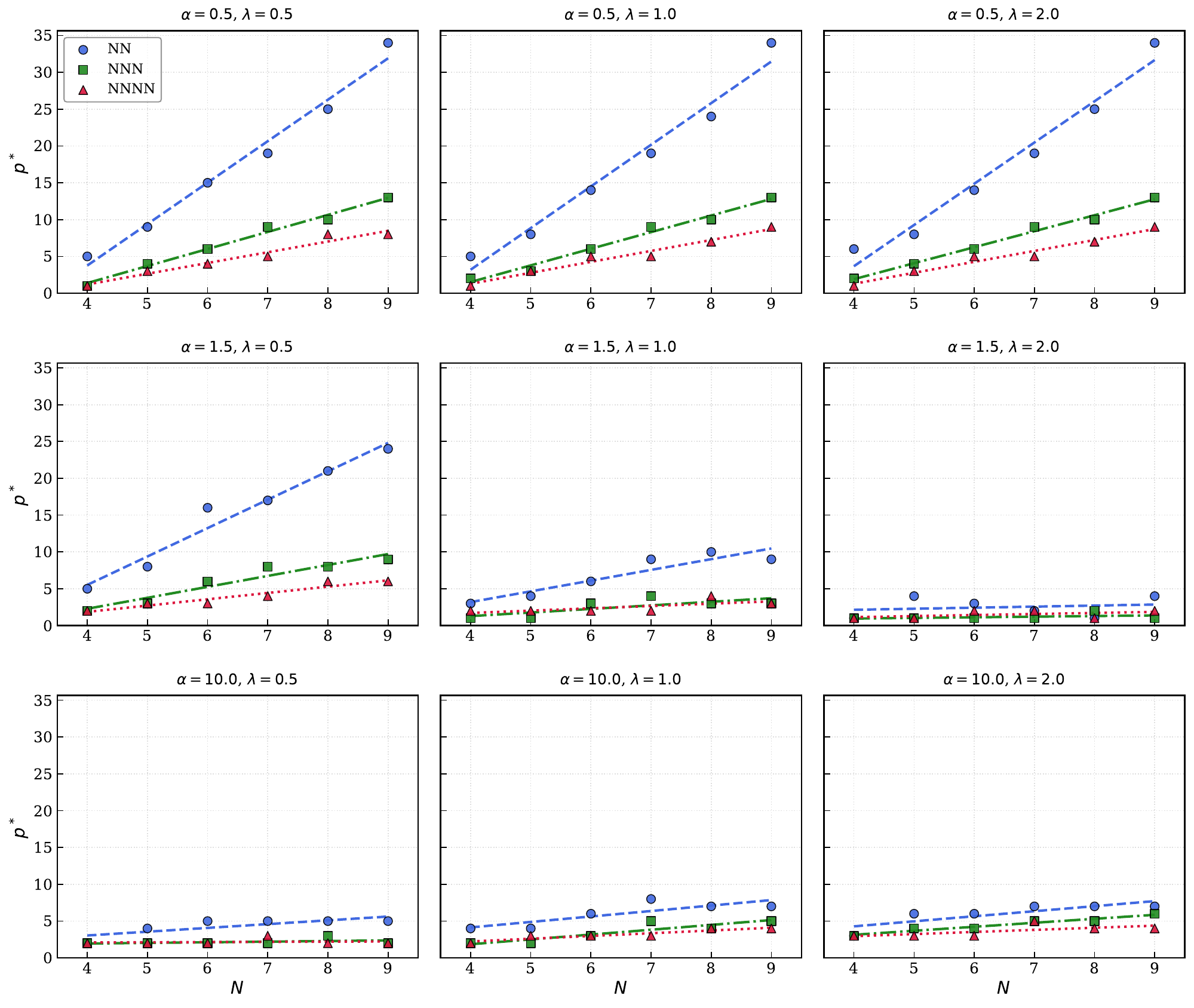}
    \caption{Minimal number of ansatz layers $p^*$ as a function of system size $N$ for the three ans\"{a}tze (NN, NNN, NNNN) across all three interaction regimes ($\alpha = 0.5$, $1.5$, $10.0$, rows) 
    and coupling ratios ($\lambda = 0.5$, $1.0$, $2.0$, columns). Data points are shown as filled markers and linear fits $p^* = aN + b$ are overlaid as dashed curves.}
    \label{fig:pstar-vs-N}
\end{figure*}

The minimal number of layers $p^*$ is defined as the smallest number of ansatz layers required to reduce the total entanglement error $\mathcal{E}(p)$ below $10^{-3}$, which is our primary diagnostic for the role of multipartite entanglement in the ansatz. Fig.~\ref{fig:pstar-vs-N} shows $p^*$ as a function of system size $N$ for all three ans\"{a}tze across all three interaction regimes and coupling ratios, with linear fits $p^* = aN + b$ overlaid on the data.

\vspace{1mm}

\textbf{Long-range regime ($\alpha = 0.5$).}
The most striking results appear in this regime, where the ground state carries algebraically decaying correlations spanning the full chain. The NN ansatz requires a rapidly growing number of layers to capture these long-range correlations. Linear fits to the $p^*$ versus $N$ data show that the NN ansatz requires approximately $5.6$ additional layers per extra qubit, consistent across all three values of $\lambda$.

This linear growth has a clear physical explanation. The NN ansatz can only directly entangle adjacent sites within each layer. To build a correlation between two sites separated by distance $r$, entanglement must be propagated step by step through the chain, requiring at least $r$ layers. In the long-range regime, the ground state contains correlations at all distances from $1$ up to $N$. The most distant pair sets the minimum layer requirement, which scales as $O(N)$. This is precisely what the linear fits confirm. This behavior is directly connected to the information propagation structure of long-range systems: as established by Hauke and Tagliacozzo \cite{ref10}, in the long-range regime there is no effective Lieb-Robinson light cone, and correlations between distant sites are established essentially instantaneously by the Hamiltonian. A nearest-neighbor ansatz is fundamentally mismatched to this structure, constrained to propagate correlations locally regardless of how non-local the target state is.

Incorporating three-body NNN interactions substantially reduces this scaling. The fitted slope for NNN is approximately $2.2$ additional layers per extra qubit, giving a ratio of $2.6\times$ relative to NN, consistent across all three values of $\lambda$. The NNNN ansatz achieves a fitted slope of approximately $1.5$ additional layers per extra qubit, a ratio of $3.8\times$ relative to NN. These slope ratios grow slightly with $N$, indicating that the advantage of the multipartite ans\"{a}tze strengthens as the system grows larger. The reduction in layers reflects structural alignment between the circuit and the Hamiltonian: the interaction terms in the long-range extended Ising model are string operators $\sigma_n^x\prod_l\sigma_l^z\sigma_{n+r}^x$ connecting sites through intermediate $\sigma^z$ operators. The NNN entangling block $e^{-i\theta\sigma_j^x\sigma_{j+1}^z\sigma_{j+2}^x}$ implements this structure directly, establishing correlations at distance two in a single layer rather than requiring two NN layers. This is the quantum circuit analog of the classical result in Lakkaraju et al.\ \cite{ref29}: a Hamiltonian with appropriately chosen finite-range interactions can reproduce the entanglement of the fully connected long-range model, and the minimum number of interactions required grows with $\alpha$ in the same way that $p^*$ does here. The key extension beyond the classical case is that we quantify the resource cost of this mimicry on a quantum circuit, both the gate cost and the classical optimization burden, a question with direct practical consequences for NISQ hardware deployment.

A physically non-trivial feature of this regime is the insensitivity of the fitted slopes to the coupling ratio $\lambda$. Whether the system is in the paramagnetic ($\lambda = 0.5$), critical ($\lambda = 1.0$), or ferromagnetic ($\lambda = 2.0$) phase, the slope of $p^*$ versus $N$ remains essentially unchanged for each ansatz. The number of ansatz layers required is determined by how many distinct correlation distances must be established, not by the magnitude of individual correlations. Since the range over which entanglement is non-zero is set by $\alpha$ and remains unchanged across the phase diagram in the long-range regime \cite{ref7, ref8, ref29}, $p^*$ is insensitive to $\lambda$. This contradicts the standard condensed matter intuition that quantum critical points are where simulation is hardest \cite{ref5} in short-range models. Criticality is special because it is the only point where correlations become long-range, but in the long-range regime studied here, correlations are system-spanning throughout the phase diagram \cite{ref7, ref8}. The interaction range parameter $\alpha$ is therefore a more reliable predictor of the requirements for ansatz layers than the system's phase.

\vspace{1mm}

\textbf{Quasi-local regime ($\alpha = 1.5$).}
Here, the behavior of $p^*$ depends sensitively on $\lambda$, in contrast to the long-range case. The fitted slopes are substantially lower than in the long-range regime. For NN at $\lambda = 0.5$, the slope is approximately $3.9$ additional layers per extra qubit, while NNN and NNNN achieve slopes of approximately $1.5$ and $0.9$ respectively, an advantage most pronounced near the critical point. Deep in the ferromagnetic phase ($\lambda = 2.0$), all three ans\"{a}tze converge to very shallow layers with slopes near zero, reflecting the simple entanglement structure far from criticality. This phase-sensitivity has a clear physical origin: when $1 < \alpha < 2$, correlations are genuinely short-range away from criticality \cite{ref7, ref8}. Near the critical point, the correlation length grows, and the NNN and NNNN ans\"{a}tze offer their greatest advantage, while deep in the ferromagnetic phase, the ground state has simple local entanglement that any of the three circuits reproduces efficiently.

\vspace{1mm}

\textbf{Short-range regime ($\alpha = 10.0$).}
In the effectively nearest-neighbor limit, all three ans\"{a}tze perform comparably. The fitted slopes are small, and no consistent ordering emerges; no ansatz shows a reliable layer advantage over the others. This confirms that the benefit of multipartite entanglement is physically tied to regimes in which the Hamiltonian generates nonlocal correlations. More expressive circuits do not automatically outperform simpler ones \cite{ref26, ref27}. In the short-range regime, the target state is local, and the additional complexity of NNN and NNNN blocks contributes nothing while increasing the per-layer resource cost.

\subsection{Quantum Resource Cost}

\begin{figure*}[t]
    \centering
    \includegraphics[width=0.75\linewidth]{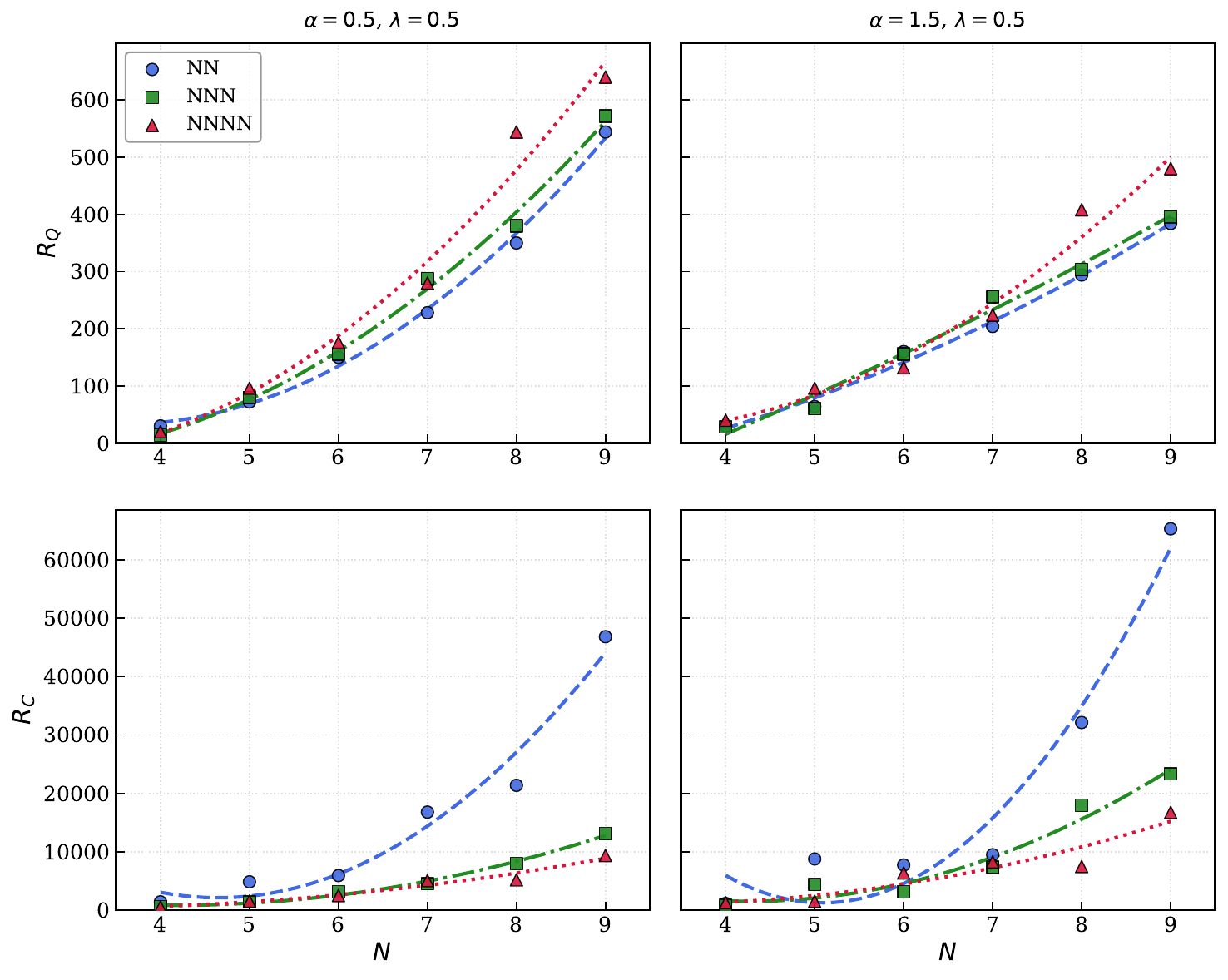}
    \caption{Quantum and classical resource costs as a function of system size $N$ for the three ans\"{a}tze (NN, NNN, NNNN) at $\lambda = 0.5$, shown for the long-range ($\alpha = 0.5$, left column) and quasi-local ($\alpha = 1.5$, right column) regimes. Top row: total quantum resource cost $R_Q$ (total CNOT gates). Bottom row: total classical resource cost $R_C$. Data points are shown as filled markers with quadratic fits overlaid as dashed curves.}
    \label{fig:rq-rc-vs-N}
\end{figure*}

The total quantum resource cost $R_Q = p^* \times R_Q^{\text{per layer}}$ accounts for both the number of layers and the CNOT gate overhead per layer. Since NNN and NNNN implement richer entangling blocks, they carry a higher per-layer gate count (Table 1): $2(3N-5)$ CNOTs per layer for NNN and $4(3N-7)$ for NNNN, compared to $2(N-1)$ for NN. At every system size and across all regimes, the total gate count follows the ordering $R_Q^{\text{NNNN}} > R_Q^{\text{NNN}} > R_Q^{\text{NN}}$, directly reflecting the higher per-layer cost of the richer entangling blocks.

However, the layer reduction of NNN and NNNN substantially narrows this gap. In the long-range regime at $N = 9$, $\alpha = 0.5$, $\lambda = 0.5$: NN requires $544$ CNOTs, NNN requires $572$, and NNNN requires $640$, differences of only $5\%$ and $18\%$ respectively, despite the per-layer CNOT counts of NNN and NNNN being approximately $3\times$ and $5\times$ that of NN at this system size. Fig.~\ref{fig:rq-rc-vs-N} (top row) shows $R_Q$ versus $N$ for $\alpha = 0.5$ and $\alpha = 1.5$ at $\lambda = 0.5$, with quadratic fits overlaid. The quadratic growth of $R_Q$ with $N$ confirms the $O(N^2)$ scaling expected from the product of a linearly growing $p^*$ and linearly growing per-layer CNOT count.

On NISQ hardware, the number of ansatz layers is often a more critical constraint than total gate count, because each additional layer requires the quantum state to remain coherent for one further sequential time step \cite{ref20}. Under this constraint, the substantially smaller $p^*$ values achieved by NNN and NNNN in the long-range and quasi-local regimes represent a practically important advantage even when their total $R_Q$ modestly exceeds that of NN. Shallower circuits accumulate less decoherence error and are more amenable to error mitigation techniques \cite{ref42}.

\subsection{Classical Resource Cost}

The classical resource cost $R_C = p^* \times L \times n_I$ captures the total optimization burden: the product of the minimal number of ansatz layers, the number of variational parameters per layer $L$, and the average number of optimizer iterations $n_I$ required to reach convergence. Since NNN and NNNN have more parameters per layer than NN (Table 1), the net $R_C$ reflects a competition between the layer reduction they offer and their larger per-layer parameter count. Fig.~\ref{fig:rq-rc-vs-N} (bottom row) shows $R_C$ as a function of $N$ for $\alpha = 0.5$ and $\alpha = 1.5$ at $\lambda = 0.5$, with quadratic fits overlaid. The fits confirm the $O(N^2)$ scaling and potentially $O(N^3)$ if $n_I$ also grows with $N$, and quantify the resource advantage more robustly than point comparisons.

In the long-range regime ($\alpha = 0.5$), NNN and NNNN consistently achieve substantially lower $R_C$ than NN across all system sizes and coupling ratios. At $\alpha = 0.5$, $\lambda = 0.5$, the leading quadratic coefficients are approximately $2{,}216$, $507$, and $221$ for NN, NNN, and NNNN, respectively, indicating reductions of $4.4\times$ and $10.0\times$ relative to NN in the large-$N$ limit. This advantage exceeds the layer reduction alone ($2.6\times$ and $3.8\times$) because both $p^*$ and $L$ are simultaneously reduced for the multipartite ans\"{a}tze, the savings in layers and parameter count compound rather than add. Since $R_C$ involves two quantities that are both linear in $N$, a constant-factor reduction in the slope of $p^*$ versus $N$ translates into a proportional reduction in $R_C$ at every system size. This multiplicative amplification grows with $N$, making structural alignment progressively more important as system size increases. The resource savings demonstrated here at $N \leq 9$ will be substantially larger at experimentally relevant system sizes.

As the interaction regime becomes more local, the advantage of the multipartite ans\"{a}tze diminishes in a physically consistent way. In the quasi-local regime ($\alpha = 1.5$), the leading quadratic coefficients at $\lambda = 0.5$ are approximately $3{,}969$, $1{,}005$, and $402$ for NN, NNN, and NNNN, respectively, still giving substantial reductions of $3.9\times$ and $9.9\times$. Away from criticality, particularly deep in the ferromagnetic phase, all ans\"{a}tze converge rapidly, and $R_C$ values are small and comparable. In the short-range regime ($\alpha = 10.0$), NN becomes competitive with or slightly below NNN and NNNN in $R_C$ for most system sizes, consistent with its lower $L$ and the already shallow $p^*$ values across all ans\"{a}tze. The added complexity of NNN and NNNN blocks contributes no layer advantage here and slightly increases the parameter count, making NN the most resource-efficient choice in this regime.

The NN ansatz additionally exhibits large variance in $R_C$ across system sizes in certain regimes. This erratic behavior originates in $n_I$: when the ansatz structure poorly matches the ground-state correlation length, the optimization landscape becomes rugged, making the optimizer sensitive to parameter initialization. NNN and NNNN exhibit more consistent $R_C$ scaling across system sizes due to their better-matched structures.

\subsection{Success Criterion: Logarithmic Negativity vs.\ Energy Fidelity}

Our finding that states with fidelity exceeding $0.99$ can fail to reproduce the correct long-range entanglement profile has a direct physical explanation. Energy is determined primarily by short-range terms in the Hamiltonian and can converge correctly even when the prepared state has incorrect quantum correlations at longer distances. Logarithmic negativity \cite{ref30, ref31}, computed via the partial transpose \cite{ref35}, probes the full spatial profile of quantum correlations and is sensitive to the algebraic decay tails that characterize the long-range regime. The entanglement error $\mathcal{E}(p)$ captures disagreement at every pairwise distance simultaneously, making it a strictly stronger convergence criterion than energy fidelity alone. We suggest that entanglement-based diagnostics should supplement energy-based criteria in any VQE study targeting long-range interacting systems. The broader principle confirmed by this work is that structural alignment between the circuit and the Hamiltonian matters more than generic expressibility \cite{ref24, ref25}. Hardware-efficient ans\"{a}tze with high expressibility \cite{ref26} do not automatically outperform structure-aware circuits - the NNN and NNNN ans\"{a}tze, which have lower generic expressibility than an all-to-all connected circuit, consistently reproduce the correct long-range entanglement at dramatically shallower layers. Expressibility is only useful when it aligns with the physics of the target state.

\section{Conclusion}
\label{sec:conclusion}
In this work, we investigated how entanglement in one layer of VQE ansatz contributes to simulate the ground state of a long-range Hamiltonian, as its interaction range varies. Taking the one-dimensional long-range extended Ising model as our work-horse, we designed three structure-aware ans\"{a}tze incorporating nearest-neighbor (NN), next-nearest-neighbor (NNN), and next-next-nearest-neighbor (NNNN) entangling blocks that directly mirror the string operators in the Hamiltonian, and evaluated performance using pairwise logarithmic negativity rather than energy fidelity alone. We show that states with fidelity exceeding $0.99$ can fail to reproduce the correct long-range entanglement profile, confirming that energy convergence alone is insufficient as a validation criterion for VQE studies of long-range systems. Pairwise logarithmic negativity, computed via partial transpose, provides a more sensitive and physically meaningful diagnostic, and we suggest that entanglement-based criteria should supplement energy-based ones in any variational study where long-range quantum correlations are physically relevant.

The central finding is that, in the non-local regime, the interaction strength $\alpha$ governs the number of ansatz layers required for quantum simulation rather than the proximity to its critical point. In the non-local regime, where the ground state carries non-local correlations throughout the phase diagram, incorporating NNN and NNNN blocks that implement the Hamiltonian's string operators reduces the layer scaling rate by factors of $2.6$ and $3.8$ relative to NN, with both quantum and classical resource costs scaling as $O(N^2)$ and the advantage growing with system size. In the quasi-local regime, the advantage of the multipartite ans\"{a}tze is strongest near the quantum critical point and fades away from criticality. In the short-range regime, all three ans\"{a}tze perform comparably, and NN is the most resource-efficient choice. The NNN ansatz provides the best overall balance across the non-local and quasi-local regimes as its layer reduction keeps the total gate cost broadly comparable to that of NN while substantially reducing the classical optimization burden, making it the most practical choice for NISQ implementation. The NNNN ansatz is most beneficial only when the number of ansatz layers is the sole binding hardware constraint. These results establish that $\alpha$ is a reliable and quantitative guide for ansatz design under NISQ constraints.

The present study was conducted on an ideal simulator without any noise. We currently have access to NISQ hardware with error-prone gates. Implementing our circuit on such NISQ hardware would lead to even shallower circuits, which in turn would severely cripple the quantum simulation. One natural way forward is to leverage error-suppression techniques \cite{ref49} to make noisy quantum simulation as efficient as possible. Moreover, our current study is limited to $N \leq 9$ qubits. Extending this analysis to larger system sizes using tensor-network based techniques \cite{ref1, ref2} is a natural next step. Adaptive ansatz methods \cite{ref25} and warm-starting strategies \cite{ref46} could also help to further reduce resource costs.


\appendix

\bibliographystyle{apsrev4-1}
\bibliography{bibliography}

\end{document}